\documentclass[fleqn,usenatbib]{mnras}
\usepackage[percent]{overpic}
\usepackage{times}
\usepackage{graphicx}
\usepackage{amsmath}
\usepackage{mathabx}
\usepackage{subfigure}
\usepackage{natbib}
\usepackage{txfonts}
\usepackage{journals}
\usepackage{xcolor}
\usepackage[utf8]{inputenc}
\usepackage[overload]{empheq}

\title[Dust in the MCs]
{Dust distributions in the Magellanic Clouds} 

\author[B.-Q. Chen et al.]
{B.-Q. Chen,$^{1}$\thanks{E-mail:
bchen@ynu.edu.cn (BQC); helong\_guo@mail.ynu.edu.cn (HLG); jiangao@bnu.edu.cn (JG).}
H.-L. Guo,$^1$\footnotemark[1]
J. Gao,$^{2}$\footnotemark[1]
M. Yang,$^{3}$
Y.-L. Liu,$^{1}$
and  B.-W. Jiang$^2$
\\
$^{1}$South-Western Institute for Astronomy Research, Yunnan University, Kunming, 650500, P.\,R.\,China\\
$^{2}$Department of Astronomy, Beijing Normal University, Beijing, 100875, P.\,R.\,China\\
$^{3}$IAASARS, National Observatory of Athens, Vas. Pavlou and I. Metaxa, Penteli 15236, Greece
}
\begin{document}

\date{Accepted ???. Received ???; in original form ???}

\pagerange{\pageref{firstpage}--\pageref{lastpage}} \pubyear{2020}
\maketitle
\label{firstpage}

\begin{abstract}
We present high-resolution maps of the dust reddening in the Magellanic Clouds (MCs). 
The maps cover the Large and Small Magellanic Cloud (LMC and SMC) area and have a spatial angular resolution between 
$\sim$ 26\,arcsec and 55\,arcmin. Based on the data from the optical and near-infrared (IR) photometric surveys, 
including the Gaia Survey, the SkyMapper Southern Survey (SMSS), the Survey of the 
Magellanic Stellar History (SMASH), the Two Micron All Sky Survey (2MASS) and the near-infrared $YJK_{\rm{S}}$ VISTA survey of the 
Magellanic Clouds system (VMC), we have obtained multi-band photometric stellar samples containing over 6 million stars in the LMC and SMC area. 
Based on the measurements of the proper motions and parallaxes of the individual stars from 
Gaia Early Data Release 3 (Gaia EDR3), we have built clean samples that contain stars 
from the LMC, SMC and Milky Way (MW), respectively. We apply the spectral energy distribution (SED) fitting to the individual sample stars to 
estimate their reddening values. As a result, we have derived the best-fitting reddening values of $\sim$ 1.9 million stars in the LMC, 
1.5 million stars in the SMC and 0.6 million stars in the MW, which are used to construct dust reddening maps in the MCs. 
Our maps are consistent with those from the literature. The resultant high-resolution dust maps in the MCs are not only 
important tools for reddening correction of sources in the MCs, but also fundamental for the studies of the distribution and properties of dust in the two galaxies.
\end{abstract}


\section{Introduction} 
\label{section1}

The Magellanic Clouds (MCs), comprising the Large and Small Magellanic Clouds (LMC and SMC), 
are the largest satellite galaxies orbiting our Milky Way (MW) galaxy. They serve as excellent astrophysical 
laboratories for the studies of the stellar populations, structure and assembly history of dwarf irregular galaxies. 
Mapping the dust distribution in the MCs are fundamental for both the reddening corrections of sources in the MCs 
and the studies of the structure and properties of the dust in the two galaxies.

In the last few decades, many attempts have been carried out to map the distribution of reddening and dust in the MCs. 
We hereby only mention a few of these works. \citet{Harris1997} presented a reddening map of the LMC based on 2,069 O 
and B main sequence stars selected from the $UBVI$ Magellanic Cloud Photometric Survey \citep{Zaritsky1997}. Based on the 
catalogue from the Magellanic Cloud Photometric Survey, \citeauthor{Zaritsky2002} (\citeyear{Zaritsky2002}, \citeyear{Zaritsky2004}) 
presented the extinction maps of the SMC and the LMC, respectively. \citet{Subramaniam2005} presented a $E(V-I)$ reddening map of the 
LMC bar based on the red clump stars selected from the Optical Gravitational Lensing 
Experiment \uppercase\expandafter{\romannumeral2} \citep[OGLE-\uppercase\expandafter{\romannumeral2};][] {Udalski2000}. 
Based on the near-IR photometric data from the Two Micron All-Sky Survey \citep[2MASS;][] {Skrutskie2006}, 
\citeauthor{Dobashi2008} (\citeyear{Dobashi2008}, \citeyear{Dobashi2009}) presented dust maps of the LMC and the SMC, respectively. 
The maps have a resolution of 2.6\,arcmin. \citet{Pejcha2009} presented a reddening map of the LMC based on $\sim$ 9000 RR Lyrae stars 
selected from the OGLE-\uppercase\expandafter{\romannumeral3} catalogue (\citealt{Soszynski2009}). 
Their map has a resolution of 0.2$^{\circ}$. \citet{Choi2018} presented a $E(g-i)$ reddening map which covers 165\,deg$^2$ of the 
LMC disk with a resolution of $\sim$ 10\,arcmin based on $\sim$ 2 million red clump stars selected from the Survey of the 
MAgellanic Stellar History \citep[SMASH;][]{Nidever2017}. \citet{Joshi2019} presented reddening maps of the 
MCs based on $\sim$ 8000 Cepheids selected from the OGLE-\uppercase\expandafter{\romannumeral4} photometric
survey (\citealt{Udalski2015}). The maps have resolutions of about 1.2\,deg$^2$ and
0.22\,deg$^2$ for the LMC and SMC, respectively. \citet{Gorski2020} presented reddening maps of the MCs with a spatial resolution 
of 3\,arcmin based on the OGLE-\uppercase\expandafter{\romannumeral3} red clump stars. \citet{Skowron2021} 
presented reddening maps of the MCs based on the OGLE-\uppercase\expandafter{\romannumeral4} red clump stars. 
The resolution of their map varies between 1.7 and 27\,arcmin.

The dust maps of the MCs show a highly inhomogeneous clumpy distribution. To accurately correct the extinction effect of objects in MCs, 
high-resolution maps are always welcome. The resolution of an extinction map is usually determined by the density of the tracer. 
Most of the previous works map the dust distributions in the MCs using a certain type of tracer in the MCs, 
such as the RR Lyrae stars (e.g. \citealt{Pejcha2009}; \citealt{Haschke2011}; \citealt{Deb2017}), Cepheids (e.g. \citealt{Inno2016}; \citealt{Joshi2019}) or 
red clump stars (e.g. \citealt{Subramaniam2005}; \citealt{Subramanian2012}; \citealt{Haschke2011}; \citealt{Tatton2013}; 
\citealt{Choi2018}; \citealt{Gorski2020}; \citealt{Skowron2021}). Their sample sizes are relatively small, which further limit 
the resolutions of the resultant reddening maps.

To study the structures and properties of the dust in the LMC and SMC themselves, one needs to consider the 
contamination of the Galactic foreground dust. As the MCs are located at high Galactic latitudes ($b$ $\sim$ $-$33\degr\ for LMC and
 $-$44\degr\ for SMC), previous works usually ignore the foreground dust contamination of the 
  MW. However, as shown in the recent three-dimensional dust map of the southern sky  \citep{Guo2021},
 there are clear Galactic foreground dust clouds toward the LMC and SMC area. The contamination of the foreground MW dust 
 should be carefully considered when studying the dust properties in the MCs.
  
In the current work, we collect data from several optical and near-infrared (IR) photometric surveys that cover the MCs, and build 
multi-band photometric stellar samples of over 6 million stars.
Based on the robust measurements of stellar proper motions and parallaxes from Gaia Early Data Release 3 (Gaia EDR3; \citealt{Gaia2021summary}), 
we have obtained clean samples that contain stars from the LMC, SMC and  MW, respectively. 
We apply spectral energy distribution (SED) fitting to the individual sample stars 
and obtained robust reddening values of about 4 million stars.
We then construct high-resolution dust reddening maps of the MCs.
The inhomogeneous Galactic foreground dust distribution in the regions toward the MCs are also mapped 
and been excluded to obtain the local dust maps of the MCs.

This paper is structured as follows. In Sect~\ref{section2}, we introduce the data used in this paper. 
In Sect~\ref{section3}, we describe our method. The results are presented in Sect~\ref{section4} and discussed in Sect.~\ref{section5}.
Finally, we summarize in  Sect~\ref{section6}.

%%%%%%%%%%  Section 2  Data  %%%%%%%%%%%%%%%%%%%%%%%
\section{Data}\label{section2}

\begin{figure*}
\centering
\includegraphics[width=0.49\textwidth]{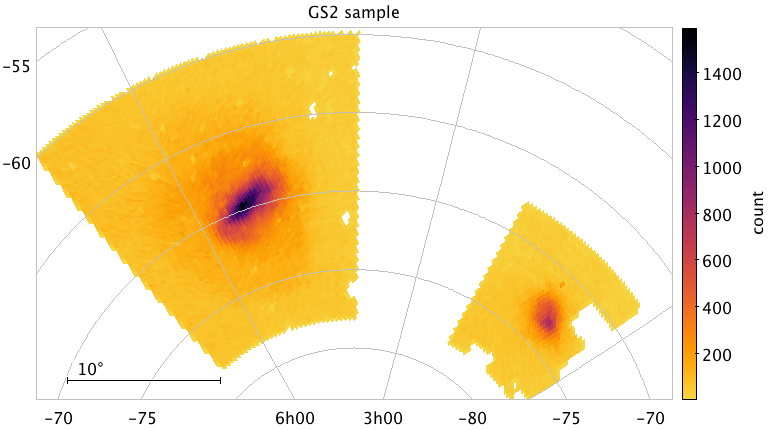}
\includegraphics[width=0.49\textwidth]{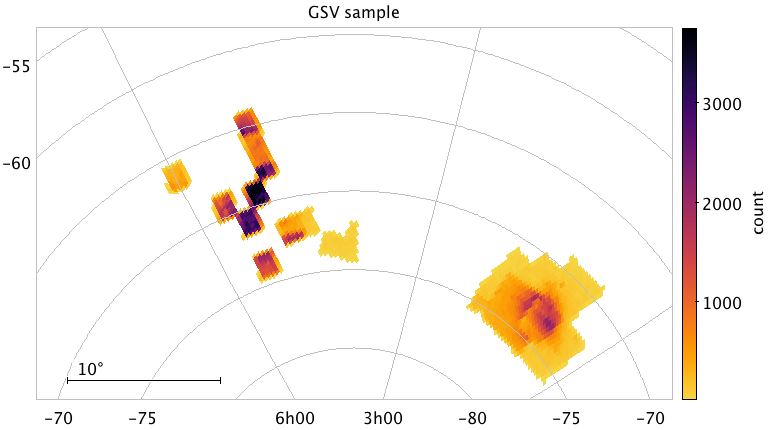}
\includegraphics[width=0.49\textwidth]{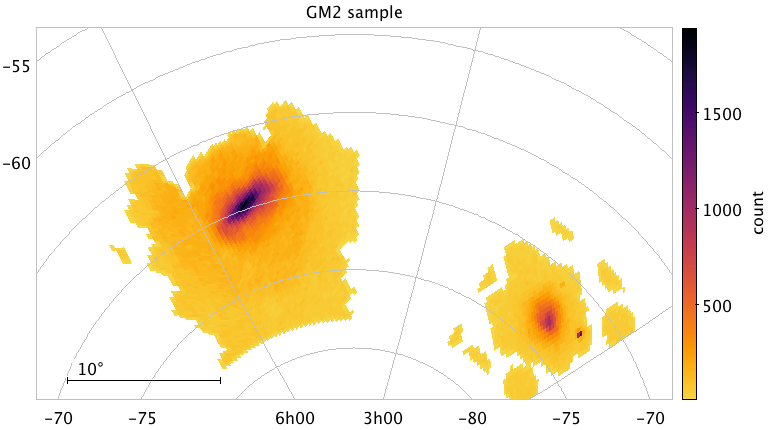}
\includegraphics[width=0.49\textwidth]{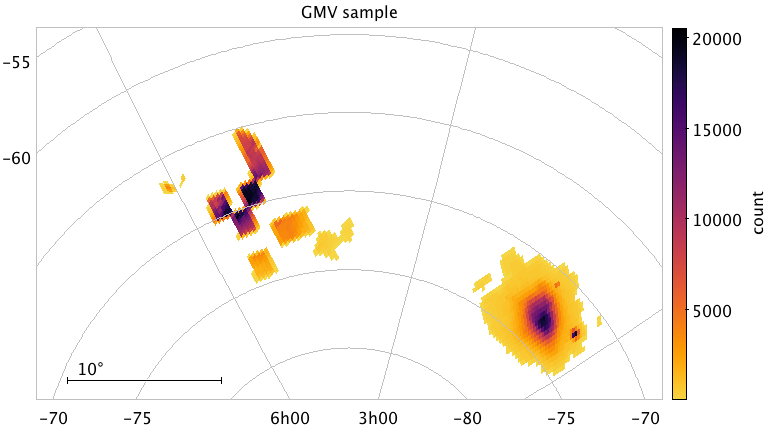}
\caption{Spatial distributions of stars from the individual catalogues. Different colours represent number of stars in bins of 0.05\,deg$^2$ size.}  
\label{spadist}
\end{figure*}

Our work is based on the optical and near-IR photometries from several multi-band photometric surveys that cover the MCs, 
including the Gaia Survey, SkyMapper Southern Survey (SMSS; \citealt{Keller2007}), SMASH, 2MASS and the near-infrared $YJK_{\rm{S}}$ 
VISTA survey of the Magellanic Clouds system (VMC; \citealt{Cioni2011}). 

The Gaia EDR3 was released in 2020. It provides high precision celestial positions and $G$ broad-band
photometric measurements for $\sim$ 1.8 billion sources. Among them, $\sim$ 1.5 billion sources have parallaxes, proper motions, the $G_{\rm{BP}}$ and $G_{\rm{RP}}$ magnitudes.
The typical uncertainties of the parallax and proper motion measurements are 0.5\,mas at $G$ = 20\,mag.
For the mean $G$-, $G_{\rm{BP}}$- and $G_{\rm{RP}}$-band photometries, the typical uncertainties are 6, 108 and 52\,mmag at $G$ = 20\,mag \citep{Gaia2021summary}.

The SMSS surveys the southern sky in six optical filters: $u,~v,~g,~r,~i$ and $z$, using the 1.35\,m SkyMapper Telescope located at Siding Spring Observatory.  
The SkyMapper Imager CCD mosaic contains 32 2048 $\times$ 4096 CCDs, with a pixel scale of $\sim$ 0.5$^{\prime\prime}$ and 
a total field-of-view of 2.4$\times$2.3\,deg$^2$. In the current work we use the SMSS Second Data Release (SMSS DR2; \citealt{Onken2019}), 
which was released in 2020. 
As measured by internal reproducibility, the photometry has a precision of 1\,per\,cent in $u$
and $v$, and 0.7\,per\,cent in $g,~r,~i$ and $z$ \citep{Onken2019}. 

The SMASH survey used the Dark Energy Camera (DECam; \citealt{Flaugher2015}) on the Blanco 4\,m Telescope at Cerro Tololo Inter-American Observatory (CTIO).
The DECam contains 62 2k$\times$4k CCDs, with a pixel scale of $0.263^{\prime \prime}$.
The SMASH survey has five optical bands: $u~,g,~r,~i$ and $z$.  
In the current work, we use the Second Data Release of SMASH (SMASH DR2; \citealt{Nidever2021SMASH}), which was released in 2020. 
The SMASH DR2 contains $\sim$ 4 billion measurements of $\sim$ 360 million objects. 
The photometry has a precision of about 1\,per\,cent in $u$ and 0.5 to 0.7\,pre\,cent in $g,~r,~i$ and $z$.

The 2MASS survey used two 1.3\,m telescopes respectively located at Mt. Hopkins and CTIO.
Each telescope is equipped with three-channel cameras using 256$\times$256 arrays of HgCdTe detectors.
2MASS has three near-IR bands: $J,~H$ and $K_{\rm{S}}$.  In the current work we use the 2MASS Point Source Catalog (2MASS PSC;  \citealt{Skrutskie2006}).
The 2MASS photometric systematic uncertainties are smaller than 0.03\,mag.  

The VMC surveyed a total area of $\sim$170\,deg$^2$ toward the MCs in three near-IR filters: $Y$, $J$ and $K_{\rm S}$, using the 4.1\,m
Visible and Infrared Survey Telescope for Astronomy (VISTA; \citealt{Emerson2004})
located at the Paranal Observatory. The 10$\sigma$ limiting magnitudes of VMC
are $\sim$ 21.1, 21.3 and 20.7\,mag respectively in the $Y,~J$ and $K_{\rm{S}}$ bands. 
In the current work we use the VMC Data Release 5.1 (VMC DR5.1; \citealt{Rubele2018}), 
which was released in 2020. The VMC DR5.1 contains 14.6 million sources
toward the SMC, the Magellanic Bridge and the Magellanic Stream. 
For the LMC area, we adopt the VMC Data Release 4 (VMC DR4; \citealt{Rubele2012}), 
which provides $YJK_{\rm S}$ photometries of sources in several fields of the LMC.

In this work, we will apply the SED fitting algorithm to the multi-band photometric data of 
the individual stars to calculate their reddening values.
To break the degeneracy between the stellar intrinsic colours and their reddening values (\citealt{Bailer2011}; \citealt{Berry2012}), we combine the optical observations 
of the individual stars (Gaia EDR3, SMSS DR2, SMASH DR2) with their near-IR data (2MASS PSC, VMC DR4 or DR5.1). 
In addition, we adopt the Gaia EDR3 astrometric measurements to distinguish the LMC or SMC stars from the MW stars. 

We first select stars located in the MCs area from the Gaia EDR3. In the current work, we define the area with right ascension (RA) between 
$4^{\rm{h}}$ and $7^{\rm{h}}$ and declination (Dec) between $-78^{\circ}$ and $-60^{\circ}$ as the LMC area, and that with RA between  
$0^{\rm{h}}$ and $2^{\rm{h}}$ and $\rm{Dec}$ between $-78^{\circ}$ and $-68^{\circ}$ as the SMC area.
We then cross-match the Gaia EDR3 stars with the optical and near-IR catalogues, respectively,
using a matching radius of 1.5\,arcsec.
We require that the sources must have detections in both the optical bands ($gri$ of SMSS DR2 or SMASH DR2) and the 
near-IR bands ($JHK_{\rm S}$ of 2MASS PSC or $YJK_{\rm S}$ of the VMC DR4 and DR5.1). As a result, we have obtained four different 
multi-band photometric samples: the Gaia/SMSS/2MASS (GS2) sample, the Gaia/SMSS/VMC (GSV) sample, 
the Gaia/SMASH/2MASS (GM2) sample, and the Gaia/SMASH/VMC (GMV) sample.
For each sample, we require that the sources must be detected in eight bands: Gaia $G_{\rm BP}G_{\rm RP}$,
SMSS/SMASH $gri$ and 2MASS $JHK_{\rm S}$/VMC $YJK_{\rm S}$, and their photometric errors in all eight bands are 
less than 0.1\,mag. These cuts lead to 845,930, 848,984, 614,648, and 4,367,440 stars in the GS2, GSV,
GM2, and GMV catalogue, respectively. For all the SMASH, SMSS, 2MASS and VMC bands, 
the errors of stars with reported photometric uncertainties less than 0.02\,mag are reset to 0.02\,mag,
which is to account for plausible calibration uncertainties. The value of 0.02\,mag comes from the systematic errors of the 2MASS
photometry \citep{Skrutskie2006}. Since we will use the empirical stellar SED libraries (Sect.~3) in the current work, we can then ignore
any effects from the zero-point calibrations of the various systems.
The spatial distributions of the individual stars in the four catalogues are shown in 
Fig.~\ref{spadist}.

\section{Method}\label{section3}

\begin{figure*}
\centering
\includegraphics[width=0.89\textwidth]{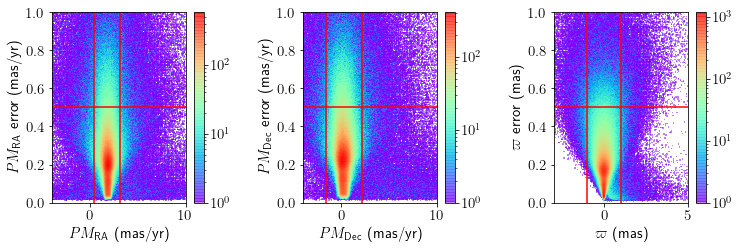}
\includegraphics[width=0.89\textwidth]{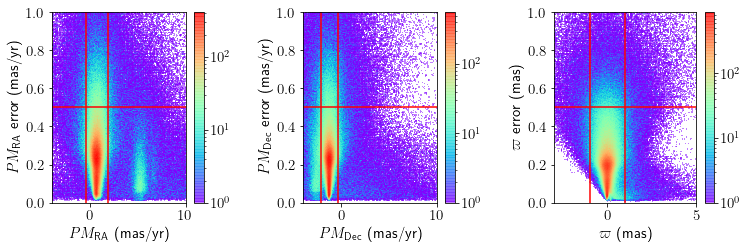}
\caption{The selection of clean LMC (upper panels) and SMC (bottom panels) source samples for the GMV catalogue. 
From left to right of each row show the Gaia EDR3 proper motions in RA, Dec and parallaxes plotted against their errors. 
The red lines show the criteria adopted to select LMC and SMC stars in the current work. }
\label{mcstarsel}
\end{figure*}

\begin{figure}
\centering
\includegraphics[width=0.49\textwidth]{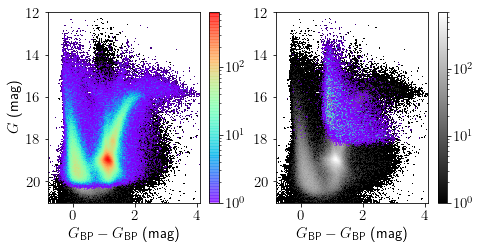}
\includegraphics[width=0.49\textwidth]{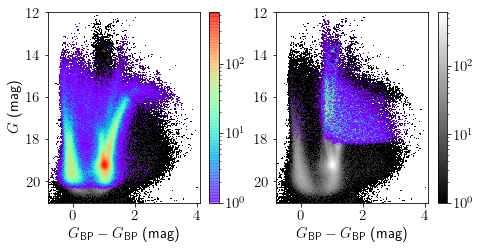}
\caption{The Gaia $G_{\rm BP} - G_{\rm RP}$ versus $G$ CMDs of of GMV catalogue stars in the LMC (upper panels) and SMC (bottom panels) regions.
The background grey images show all stars in the corresponding area. 
The colour images in the left panels show the LMC (upper) and SMC (bottom) sources 
after the astrometric selection, respectively. The colour images in the right panels show the selected MW foreground sources in the corresponding area.}
\label{mcstarcmd}
\end{figure}

\subsection{The LMC, SMC and MW samples}\label{section3.1}

We first establish clean source catalogues of stars respectively from the LMC, SMC and MW.
We adopt a method similar to that used by \citeauthor{Yang2019} (\citeyear{Yang2019}, \citeyear{Yang2021}) and \citet{Gaia2021MCs}.
The Gaia EDR3 astrometric measurements \citep{Gaia2021summary} are used to determine the membership of LMC and SMC stars.
Fig.~\ref{mcstarsel} shows the selections of the LMC and SMC stars in this work. We select only stars with parallax and proper motions within the 3$\sigma$ limits
of LMC and SMC, respectively. We adopt the mean and dispersion values of the astrometry of the LMC and SMC stars  from \citet{Gaia2021MCs}.
For LMC star selection, we adopt the criteria:
proper motion in RA $0.4192<PM_{\rm RA}<3.1024$\,mas/yr, proper motion in Dec 
$-1.6087<PM_{\rm Dec}<2.2163$\,mas/yr and parallax  $-1.0078 < \varpi < 0.9998$\,mas.
For SMC, we adopt the criteria:
$-0.3863<PM_{\rm RA}<1.8505$\,mas/yr, 
$-2.1232<PM_{\rm Dec}<-0.3280$\,mas/yr and   $-0.9845 < \varpi < 0.9793$\,mas.
In addition, we have removed stars with proper motion errors larger than 0.5\,mas/yr and parallax errors 
larger than 0.5\,mas. As a result, we have obtained a total of $\sim$ 3 million LMC stars and 1.7 million SMC  
stars from the four multi-band photometric stellar catalogues mentioned in Sect.~2.

We have also selected a clean MW star sample using the criteria: $\varpi >  0.2$\,mas, parallax error $0<\sigma_{\varpi} < 0.1$\,mas 
and the relative uncertainty of parallax $\sigma_{\varpi}/\varpi < 0.2$. This yields a total of $\sim$ 0.7 million 
MW stars in the LMC and SMC areas. We note that we have adopted very
strict constraints when we select the Galactic foreground stars. We have excluded any stars locating at distances larger than 5\,kpc
from the Sun or having large parallax errors. This is because that we want to select a Galactic foreground star sample as clean as possible. The Gaia EDR3
parallaxes of sources located at distances $d$ $>$ 5\,kpc from the Sun usually suffer larger uncertainties.
As the MCs are located at high Galactic latitudes, the dust from the MW mainly locates at close distances 
($d$ $< 1$\,kpc; \citealt{Chen2013}; \citealt{Schultheis2014}; 
\citealt{Chen2019}; \citealt{Green2019}). We can then use our selected MW stars to map the Galactic 
foreground dust distribution of the regions toward the MCs.

In Fig.~\ref{mcstarcmd} we show the Gaia $G_{\rm BP} - G_{\rm RP}$ versus $G$ colour-magnitude diagram of our selected 
LMC, SMC and MW stars from the GMV catalogue. Based on the astrometric constraints, we are able to obtain clean LMC, SMC and MW stellar samples. 

\begin{table}
\centering
\caption{The extinction coefficient for all passbands adopted in the current work.}
\begin{tabular}{ccccc}
\hline
\hline
  &   &   &   $R_{\lambda}$     &  \\
\cline{3-5}
Filter                            &     Effective wavelength (\AA)     &      MW       &          LMC            &       SMC      \\
\hline
Gaia $G_{\rm{BP}}$        &    5182.58$^a$                         &        3.324  &        3.639          &       2.960    \\
Gaia $G_{\rm{RP}}$        &    7825.08$^a$                         &        1.932  &        2.179          &       1.645    \\
SMASH  $g$                      &    4841.88$^b$                         &        3.628  &        3.943          &       3.263    \\
SMASH  $r$                      &    6438.53$^b$                         &        2.595  &        2.880          &       2.263    \\
SMASH  $i$                    &    7821.01$^b$                         &        1.933  &        2.181          &       1.646    \\
SMSS  $g$                   &    5099.44$^c$                         &        3.392  &        3.707          &       3.027    \\
SMSS  $r$                   &    6157.28$^c$                         &        2.731  &        3.025          &       2.390    \\
SMSS  $i$                   &    7778.37$^c$                         &        1.953  &        2.201          &       1.665    \\
VMC  $Y$              &    10573.79$^d$                        &        1.145  &        1.307          &       0.956     \\
VMC $J$              &    21771.24$^d$                        &        0.845  &        0.964          &       0.705     \\
VMC $K_{\rm{S}}$     &    21475.88$^d$                        &        0.366  &        0.418          &       0.305    \\
2MASS $J$                  &    12357.60$^e$                        &        0.889  &        1.015          &       0.742    \\
2MASS $H$                  &    16476.02$^e$                        &        0.561  &        0.640          &       0.468     \\
2MASS $K_{\rm{S}}$         &    21620.75$^e$                        &        0.362  &        0.413          &       0.302     \\
\hline
\end{tabular}
    \parbox{\textwidth}{\footnotesize \baselineskip 3.8mm
Note: $^a$\citet{Riello2021}, $^b$\citet{Abbott2018}, 
$^c$\citet{Bessell2011}, \\
$^d$\citet{Dalton2006}, $^e$\citet{Cohen2003}.}
\label{Table1}
\end{table}

\begin{figure}
\centering
\includegraphics[width=0.49\textwidth]{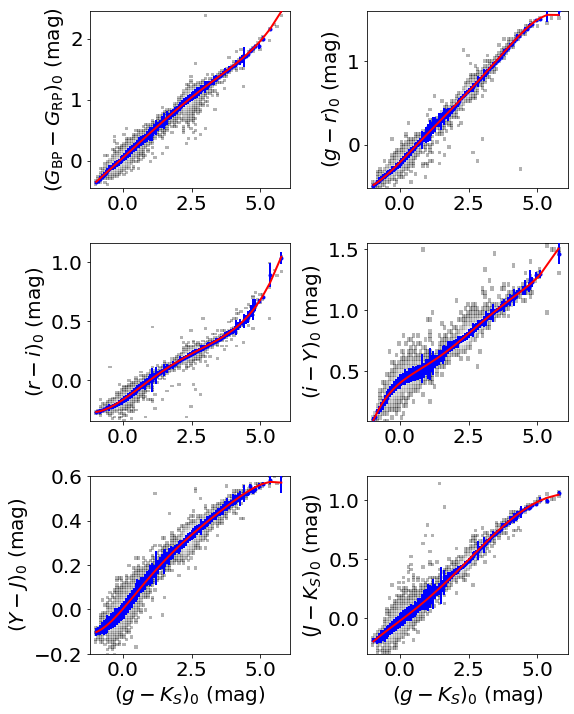}
\caption{
Colour–colour diagrams of all stars from the LMC reference sample of the GMV catalogue.
The background grey contours are on a logarithmic scale. Blue points and error bars represent the
median values and standard deviations of the colours for the individual $(g-K_{\rm S})_0$ bins. 
The Red lines are the fifth polynomial fits to the median values.
}
\label{sllmc}
\end{figure}

\begin{figure}
\centering
\includegraphics[width=0.49\textwidth]{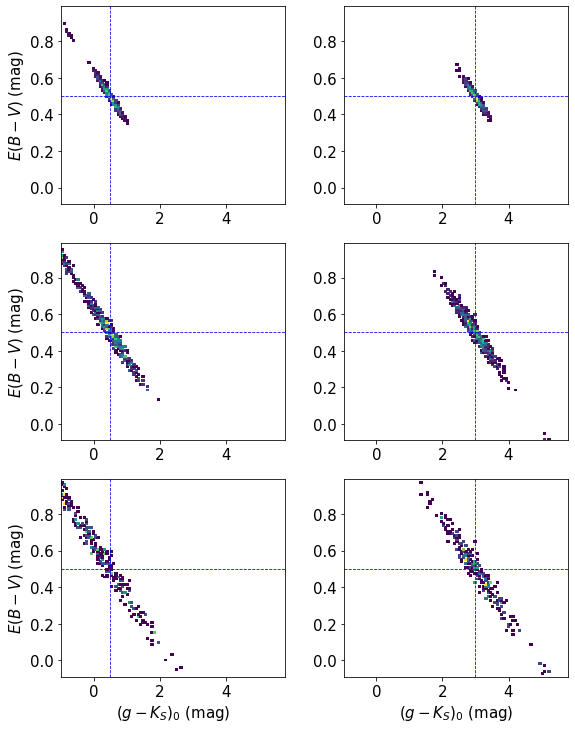}
\caption{
Monte Carlo simulations for two GMV fiducial stars in the LMC of intrinsic colours $(g - K_S)_0$ = 0.5 (left panels) 
and 3.0\,mag (right panels).
Each panel shows the distribution of the best-fit results in the $(g - K_S)_0$ and $E(B-V)$ plane. 
The upper, middle and bottom rows refer to the results of the assumed
photometric errors of 0.02, 0.05 and 0.08\,mag, respectively. The two dashed lines in each panel mark the assumed
$(g - K_S)_0$ and $E(B-V)$ values of the stars.
}
\label{fidtest}
\end{figure}

\begin{figure*}
\centering
\includegraphics[width=0.36\textwidth]{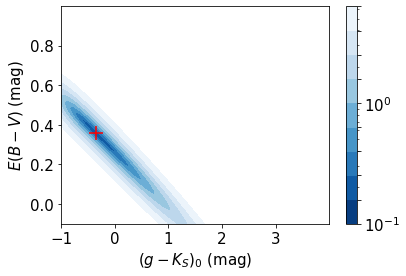}
\includegraphics[width=0.36\textwidth]{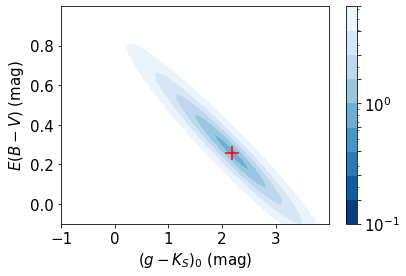}\\
\includegraphics[width=0.36\textwidth]{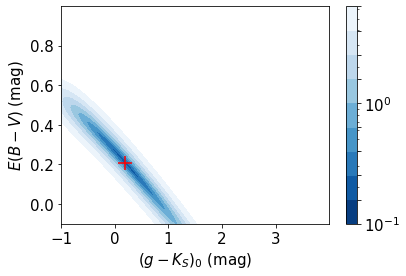}
\includegraphics[width=0.36\textwidth]{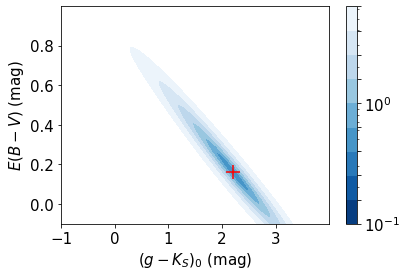}
\caption{The  $\chi ^2$ surfaces for four example stars in the GMV catalogue
(upper left: a blue star in the LMC, upper right: a red star in the LMC, bottom left: a blue star in SMC, 
and bottom right: a red star in the SMC). For all panels, the best-fitted $(g-K_{\rm S})_0$ and $E(B-V)$ 
values are marked with red pluses.}  
\label{gkebv}
\end{figure*}

\begin{table}
\centering
\caption{Parameters of the reference samples.}
\begin{tabular}{ccccc}
\hline
\hline
 Sample & $E(B-V)_{\rm max}$ & N   &   $(g-K_{\rm S})_{0,{\rm min}}$   &  $(g-K_{\rm S})_{0,{\rm max}}$  \\
  & mag & & mag & mag \\
\hline
GS2-LMC  &  0.050   & 14,219 & $-$0.2  & 5.7 \\
GS2-SMC  &  0.055  & 9,500   & $-$0.8   & 7.2 \\
GS2-MW   & 0.035   & 38,977  & $-$0.1 & 6.2 \\
GSV-LMC  &  0.035  & 10,001 & $-$1.0   & 5.3 \\
GSV-SMC  &  0.035  & 15,949 & $-$1.0  & 6.2 \\
GSV-MW   & 0.035    & 13,278 & 0.5      & 5.8 \\
GM2-LMC  & 0.065   & 9,739   & 0.5      & 5.7 \\
GM2-SMC  &  0.055  & 9,305   & 1.2      &  7.8 \\
GM2-MW    &  0.045 & 12,831  & 0.8      & 6.2 \\
GMV-LMC  & 0.055   &  14,504 & $-$1.0 & 5.8 \\
GMV-SMC  &  0.035  & 124,802 & $-$1.0    & 8.0 \\
GMV-MW   &  0.030   & 12,415  & $-$1.0    & 6.7 \\
\hline
\end{tabular}
\label{sl1}
\end{table}

\begin{figure*}
\centering
\includegraphics[width=0.99\textwidth]{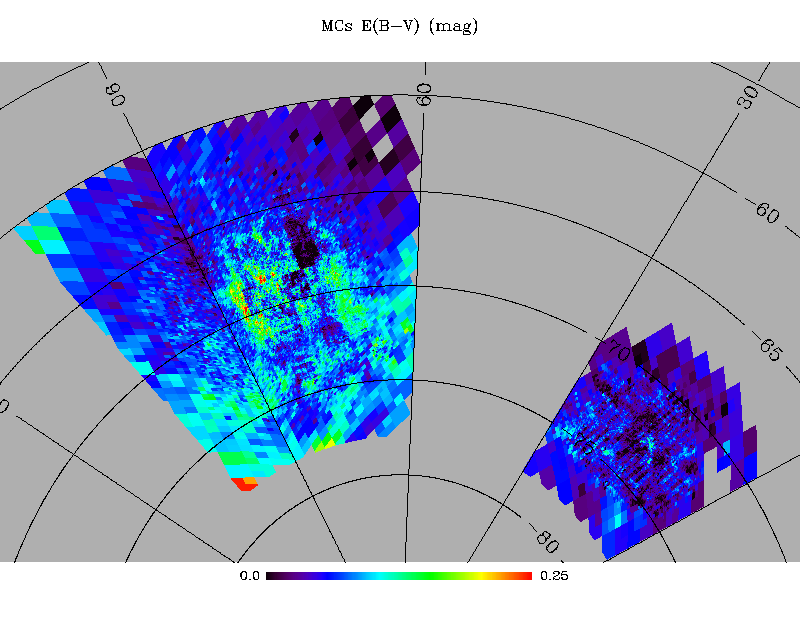} \\
\includegraphics[width=0.35\textwidth]{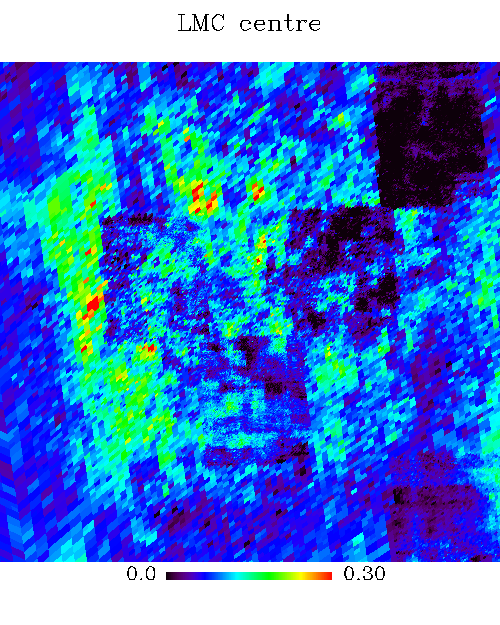}
 \hspace{4.5cm}
\includegraphics[width=0.35\textwidth]{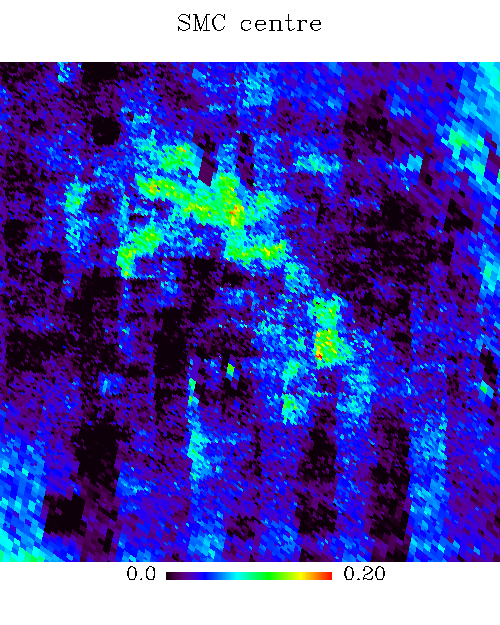}
\caption{The reddening maps of the MCs. The bottom panels show the zoomed maps toward the LMC (left) and the 
SMC centre (right), respectively.}
\label{ebv2d}
\end{figure*}

\begin{figure*}
\centering
\includegraphics[width=0.99\textwidth]{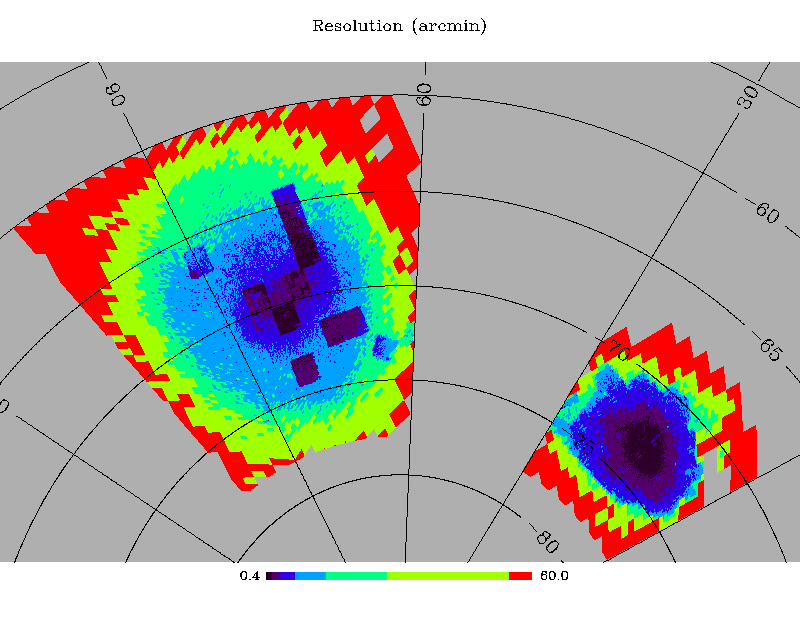} \\
\caption{The angular resolution of our reddening maps.}
\label{reso}
\end{figure*}

\begin{figure}
\centering
\includegraphics[width=0.35\textwidth]{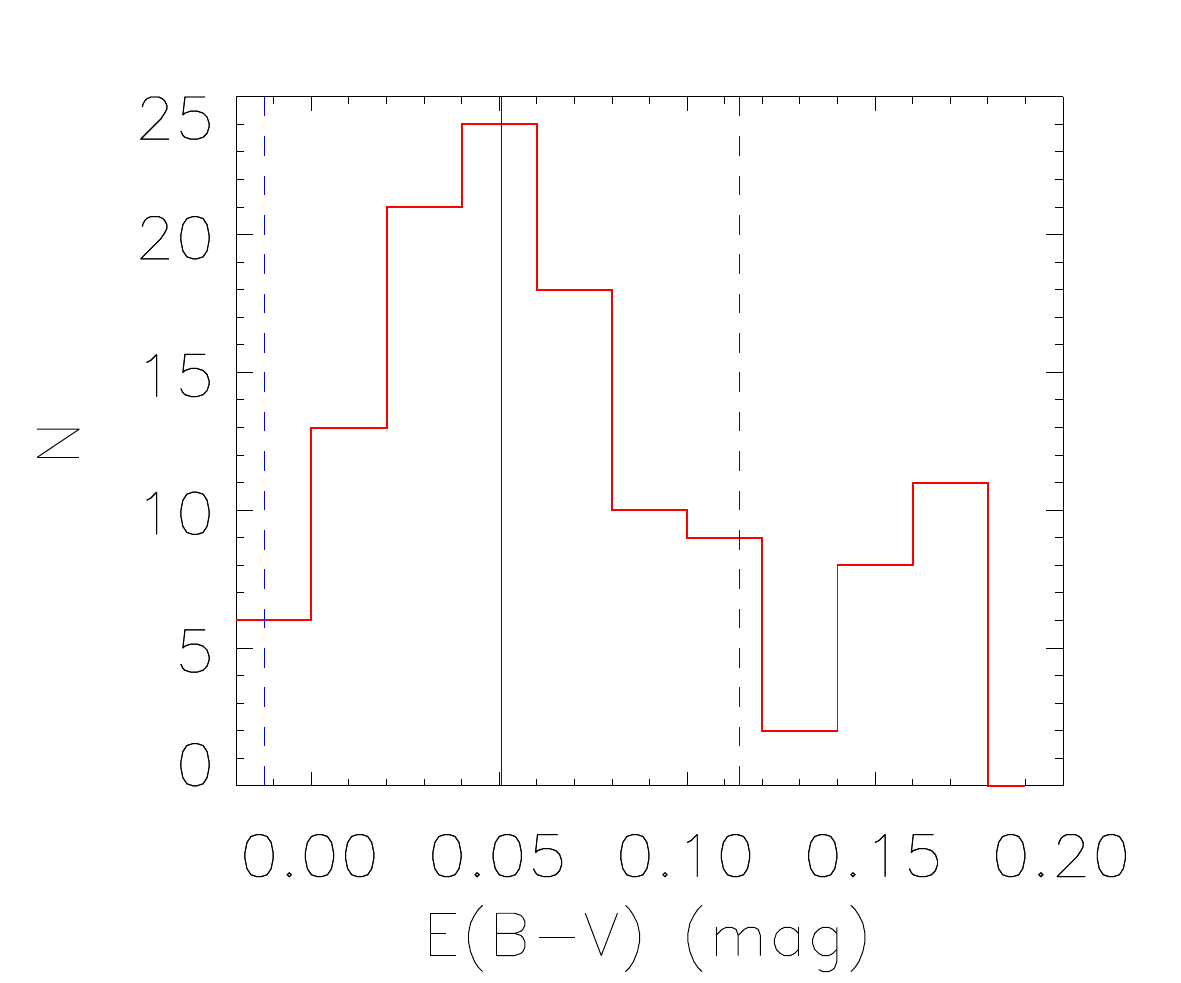}
\caption{The distribution of reddening values for all stars in an example subfield (pixel) of the SMC. 
The vertical blue solid and dashed lines represent our resultant reddening value and the reddening uncertainty of the pixel.}  
\label{ebvdist}
\end{figure}

\begin{figure*}
\centering
\includegraphics[width=0.99\textwidth]{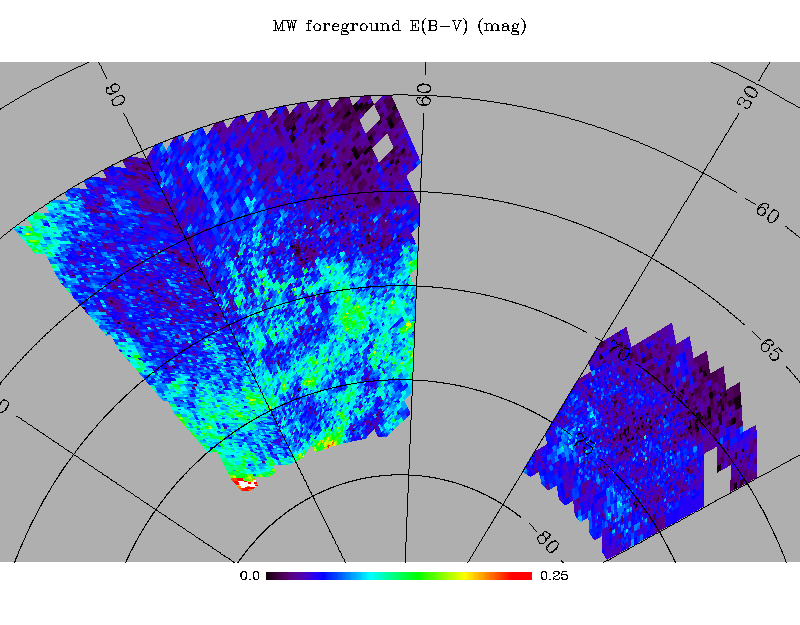} \\
\caption{The MW foreground reddening maps of the sky area toward the MCs.}
\label{mwebv}
\end{figure*}

\begin{figure*}
\centering
\includegraphics[width=0.99\textwidth]{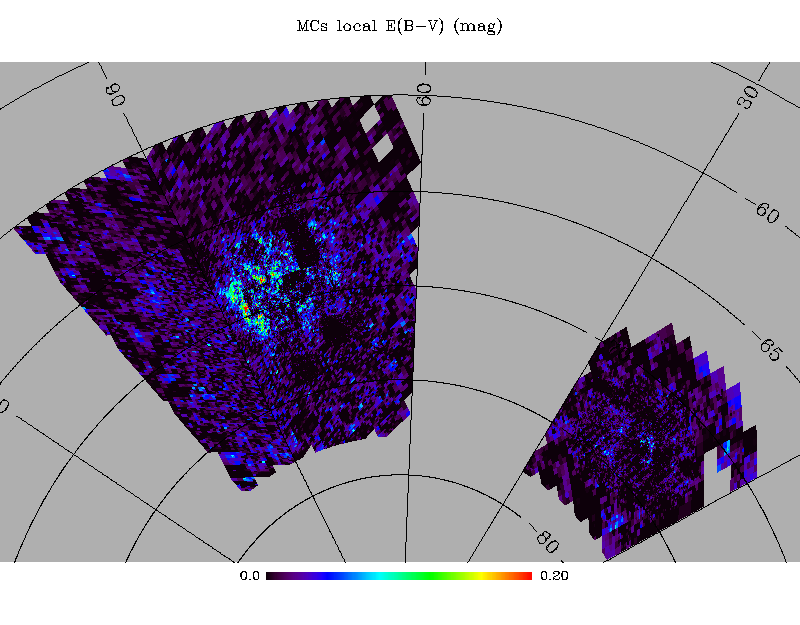} \\
\includegraphics[width=0.35\textwidth]{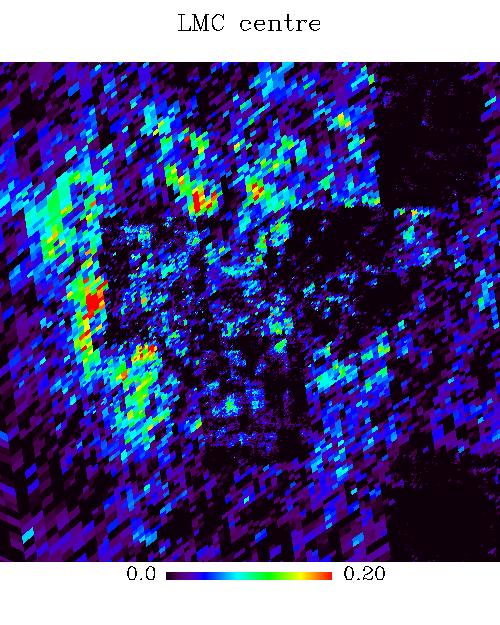}
 \hspace{4.5cm}
\includegraphics[width=0.35\textwidth]{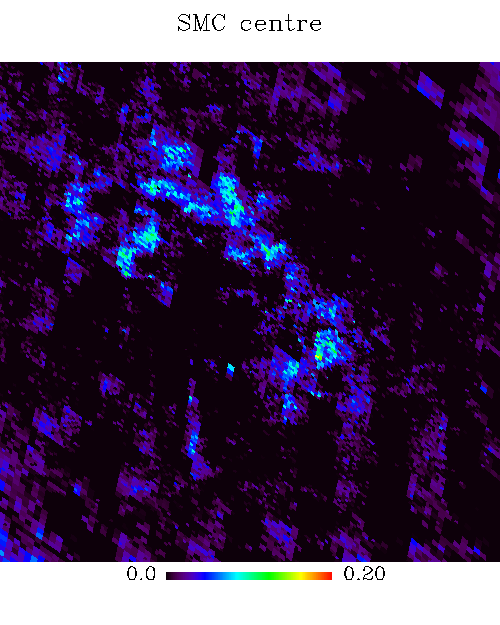}
\caption{The local dust distributions of the LMC and SMC galaxies.
The bottom panels show the zoomed local reddening maps of the LMC (left) and the 
SMC centre (right), respectively.}
\label{mclocal}
\end{figure*}

\subsection{Reddening determinations for the individual stars}\label{section3.2}
In this work, we adopt a SED fitting algorithm similar to that of \citet{Berry2012}, \citet{Chen2014} and \citet{Guo2021} 
to the multi-band photometric data of the individual stars to calculate their reddening values. For a given star, 
we simulate its observed colour of bands $\lambda_1$ and $\lambda_2$ by,
\begin{equation}\label{equation1}
c_{\rm{sim}} = c_{\rm{sl}}[(g-K_{\rm{S}})_0] + (R_{\lambda_1} - R_{\lambda_2}) \cdot E(B-V),
\end{equation}
where the intrinsic color $(g-K_{\rm{S}})_{\rm{0}}$ and reddening value $E(B-V)$ are free parameters to fit. 
$c_{\rm{sim}}$ is the simulated color of bands $\lambda_1$ and $\lambda_2$ for the star and $c_{\rm{sl}}[(g-K_{\rm{S}})_0]$ the 
corresponding intrinsic color predicted by the reference stellar locus for given $(g-K_{\rm{S}})_{\rm{0}}$. 
$R_{\rm{\lambda_1}}$ and $R_{\rm{\lambda_2}}$ are the extinction coefficients of bands $\lambda_1$ and $\lambda_2$, respectively.
For the extinction law, we adopt the results from \citet{Gordon2003}. 
We assume $R_V$ = 3.41, 2.74 and 3.1 for the LMC, SMC and MW respectively. The \citet*{CCM1989} $R_V$-dependent 
extinction curves are adopted to obtain the extinction coefficients of the individual passbands, which are listed in Table\,\ref{Table1}.

We have calculated the empirical stellar locus for our selected GS2, GSV, GM2 and GMV multiband catalogues, respectively. For each catalogue, we select reference samples respectively for the LMC, SMC and MW stars
to obtain their stellar locus. We require that reference sample stars have photometric errors in all bands smaller than 0.05\,mag and 
have low line-of-sight extinction from the extinction map of \citet*[SFD hereafter]{SFD1998}. 
The upper limits of the extinction vary from $E(B-V)_{\rm max}$ = 0.030\,mag
to 0.055\,mag for the individual reference samples to obtain enough stars.
In Table~\ref{sl1} we list the extinction upper limit $E(B-V)_{\rm max}$ and the star counts $N$ of all the reference samples.
The reddening effects of the reference stars are corrected using the extinction values from the SFD map and the reddening 
coefficients from Table~\ref{Table1}. 
In the current work, we assume the intrinsic colour $(g-K_{\rm S})_0$ as the independent variable and use 
fifth-order polynomials to fit the running medians of the individual colours to obtain the stellar locus.
The  lower and upper $(g-K_{\rm S})_0$ limits [$(g-K_{\rm S})_{0,{\rm min}}$  and  $(g-K_{\rm S})_{0,{\rm max}}$ )]
of all the reference samples are also presented in Table~\ref{sl1}.
As an example, we show the example stellar locus resulted from
the LMC reference sample of the GMV catalogue in Fig.~\ref{sllmc}. 

The best-fit intrinsic color $(g-K_{\rm{S}})_0$ and reddening value $E(B-V)$ of a given star are found by minimizing $\chi^2$ defined as
\begin{equation}\label{equation2}
\chi^2=\frac{1}{4}\sum_{i=1}^{6}(\frac{c_{\rm{obs}}^i-c_{\rm{sim}}^i}{\sigma_i})^2,
\end{equation}
where $c^{i}_{\rm{obs}}$ are 6 observed colors ($G_{\rm{BP}}-G_{\rm{RP}}$, $g-r$, $r-i$, $i-J$, $J-H$ and $H-K_{\rm{S}}$ for the GS2 and GM2 catalogues,
and $G_{\rm{BP}}-G_{\rm{RP}}$, $g-r$, $r-i$, $i-Y$, $Y-J$ and $J-K_{\rm{S}}$ for the GSV and GMV catalogues), 
$c^{i}_{\rm{sim}}$ and $\sigma_i$ are the corresponding simulated colours and colour uncertainties, respectively. 
The colour uncertainties $\sigma_i$ are the combination of the photometric uncertainties and stellar locus fitting errors. 

The optimization is carried out by running a pseudo-$E(B-V)$ ranging from $-$0.1\,mag to $E(B-V)_{\rm{SFD}} + 0.1$\,mag in step of 0.002\,mag 
and $(g-K_{\rm{S}})_0$ ranging from $(g-K_{\rm S})_{0,{\rm min}}$
to $(g-K_{\rm{S}})_{\rm{obs}} + 0.1$\,mag in step of 0.01\,mag, 
where $E(B-V)_{\rm{SFD}}$ is the reddening value of a star from the SFD map and 
$(g-K_{\rm{S}})_{\rm{obs}}$ is the observed colour values. If the value of $(g-K_{\rm{S}})_{\rm{obs}} + 0.1$ is larger than 
$(g-K_{\rm S})_{0,{\rm max}}$, we will adopt $(g-K_{\rm S})_{0,{\rm max}}$ 
as the maximum value of the pseudo-$(g-K_{\rm{S}})_0$. As the reddening values in the MCs area are relatively low, 
we have set a negative lower limit of the reddening values to avoid the systematically overestimation of reddening values for regions of 
low extinction (\citealt{Berry2012}; \citealt{Schlafly2014}; \citealt{Chen2014,Chen2015}).

As shown in Fig.~9 of \citet{Berry2012}, the intrinsic colour and reddening degeneracy 
can be broken when we combine the optical and the near-IR colours. However, due to the photometric errors 
and the fact that the stellar locus orientation in the near-IR bands is not perpendicular to the reddening vectors, 
the covariance between the best-fit intrinsic colours $(g-K_s)_0$ and reddening $E(B-V)$ values does not vanish. 
We have simulated the GMV colours of fiducial stars in the LMC based on our derived stellar locus. We assume 
different intrinsic colours $(g - K_S)_0$ = 0.5 and 3.0\,mag, representing blue and red stars, respectively. 
We set the reddening value $E(B-V)$ = 0.5\,mag. By adopting three sets of photometric errors, 0.02, 0.05 and 0.08\,mag, 
the `observed' colours of the fiducial stars are generated. For each star, we randomly generate their colours
and calculate the best-fit parameters $(g - K_S)_0$ and $E(B-V)$ by our SED fitting algorithm 1000 times. 
The distributions of the best-fit parameters are shown in Fig.~\ref{fidtest}. 
Overall, our algorithm is able to recover robust values of $(g-K_S)_0$ and $E(B-V)$ for both the blue and stars.
Similar as in the work of \citet{Berry2012} and \citet{Chen2014}, the $(g - K_S)_0$ 
and $E(B-V)$ covariance is larger for a blue star than a red one. For a star with larger photometric errors, 
the covariance is larger. For the three assumed cases of photometric errors, 0.02, 0.05 and 0.08\,mag, the 
dispersions of reddening $E(B-V)$ values are 0.09, 0.21 and 0.26\,mag, respectively, for the blue star, and 
0.05, 0.16 and 0.23\,mag, respectively, for the red star.

We have also investigated the covariance between the stellar intrinsic 
colour $(g-K_{\rm{S}})_0$ and the reddening $E(B-V)$ values for real stars in our sample.
The resultant $\chi ^2$ surfaces for four example blue and red stars in both the LMC and SMC samples are illustrated in Fig.~\ref{gkebv}. 
The covariance between the stellar intrinsic colour $(g-K_{\rm S})_0$ and the reddening $E(B-V)$ values of the stars in the MCs is very similar 
to that of the Galactic stars from the tests in \citet[ their Fig.~12]{Berry2012} and \citet[ their Fig.~5]{Chen2014}. 
For all the four example stars, the $\chi ^2$ distributions always possess only one peak.
The $E(B-V)$ versus $(g-K_{\rm S})_0$ covariance of the stars in the LMC is very similar to that of the stars in the SMC, 
and it does not strongly depend on the reddening $E(B-V)$ values. 
These features indicate that our $\chi ^2$ minimization SED fitting  
algorithm produces statistically correct results. With the combination of the optical and near-IR photometries, 
we are able to break the degeneracy between the intrinsic stellar colour and the dust reddening.

\subsection{Construction of high-resolution reddening maps}\label{section3.5}

Based on the reddening estimates of the individual stars, we are then able to map the dust reddening distributions of the MCs. 
We use the HEALPix pixelization scheme \citep{Gorski2005} as our method to divide the sky into individual subfields (pixels).
In the current work, we have adopted a variable resolution based on stellar density.  
We require that there are at least 20 stars in each pixel. The resolution of the pixels varies from $\sim$ 26\,arcsec 
(with HEALPix nside $=$ 8,192) to $\sim$ 55\,arcmin (with HEALPix nside $=$ 64).
For each pixel, we first exclude outliers that have reddening values outside the 3$\sigma$ limit. 
The average and variance of the reddening values of the remaining stars are respectively 
adopted as the reddening value and the reddening uncertainty
of the pixel.

\begin{figure}
\centering
\includegraphics[width=0.45\textwidth]{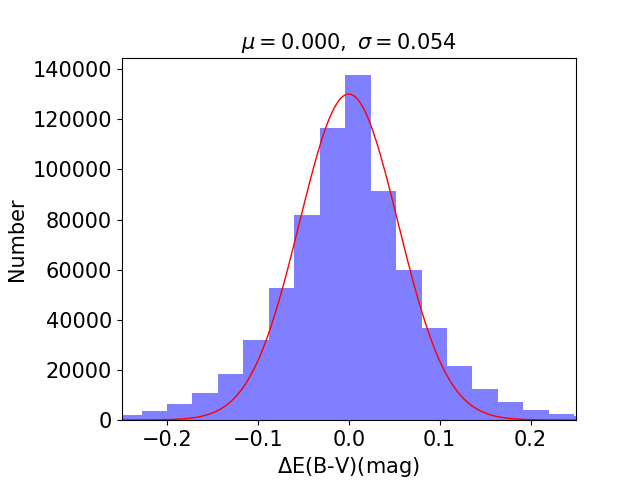}
\caption{Histogram of differences of reddening values by 
common stars derived from different multiband photometric samples
in the current work. A Gaussian fit to the distribution is overplotted, and the mean and dispersion
of the Gaussian are marked in the diagram.}
\label{intern}
\end{figure}

\begin{figure}
\centering
\includegraphics[width=0.45\textwidth]{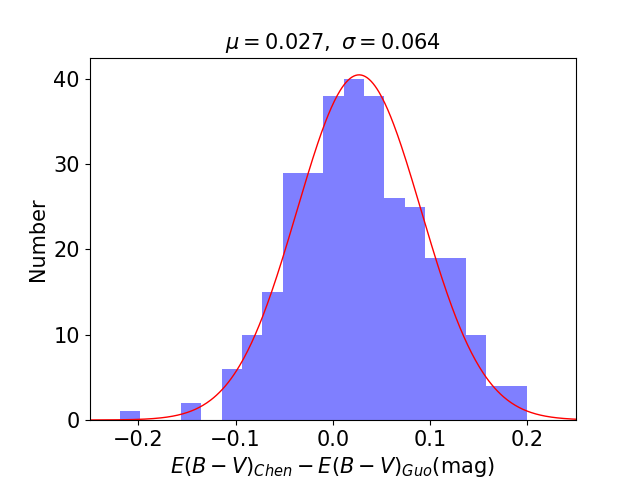}
\caption{Histograms of differences of reddening values of the individual stars
derived from the current work and those given by \citet{Guo2021}. 
A Gaussian fit to the distribution is overplotted, and the mean and dispersion
of the Gaussian are marked in the diagram.]}
\label{guo20}
\end{figure}

\begin{figure*}
\centering
\includegraphics[width=0.45\textwidth]{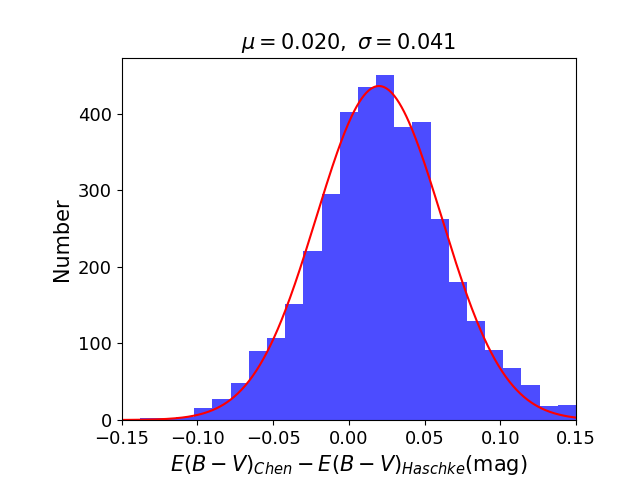}
\includegraphics[width=0.45\textwidth]{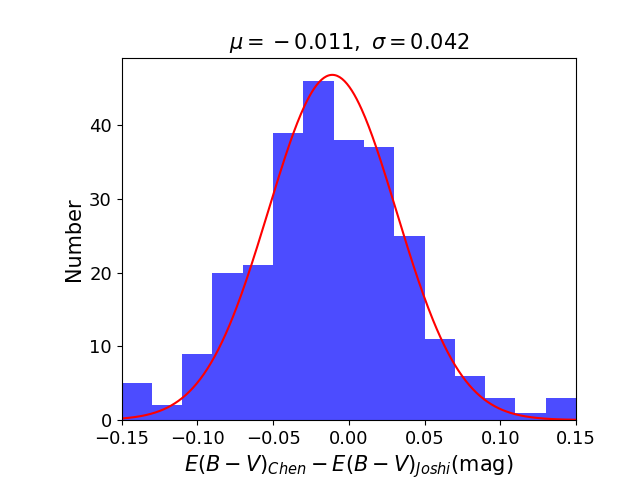}
\includegraphics[width=0.45\textwidth]{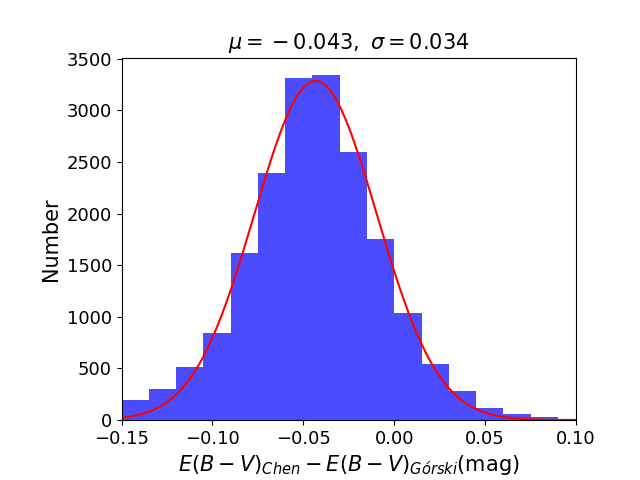}
\includegraphics[width=0.45\textwidth]{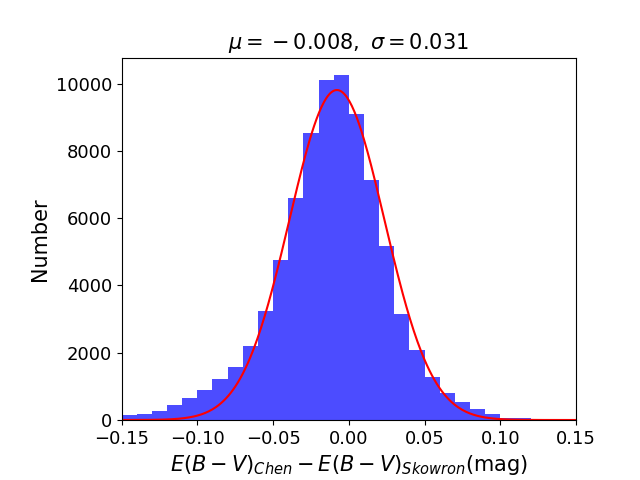}
\caption{Histograms of differences of the reddening values of our reddening maps and those from
\citet{Haschke2011}, \citet{Joshi2019}, \citet{Gorski2020} and  \citet{Skowron2021}. 
Gaussian fits to the distributions are overplotted, 
and the mean and dispersion of the Gaussian are marked in each panel.}
\label{litera}
\end{figure*}

\section{Result}\label{section4}

We have applied our SED fitting algorithm to the individual sample stars. In the current work, we exclude the bad
SED fits with a minimum $\chi^2$ larger than 2. This yields a total number of 4,037,497 stars in our final catalogue, including 
$\sim$ 1.9 million LMC stars, 1.5 million SMC stars and 0.6 million MW stars. 

\subsection{Reddening maps of the MCs}\label{section4.1}

The LMC and SMC stars in our final catalogue are used to construct the high resolution reddening maps of the MCs, 
which are plotted in Fig.~\ref{ebv2d}.  Our reddening maps cover about 360\,deg$^2$ sky area toward the MCs. 
The average reddening value is $E(B-V)$ $=$ 0.07\,mag in the LMC area and 
$E(B-V)$ $=$ 0.04\,mag in the SMC area. In the LMC area, the reddening values increase and exceed $E(B-V)$ = 0.38\,mag  
in the region of the famous CO arc cloud (RA $\sim$ 87.0\degr\ and Dec $\sim$ $-$70.1\degr; \citealt{Dobashi2008}).
In the SMC area, the reddening values increase and reach its maximum value with $E(B-V)$ = 0.2\,mag in the region of the 
SMC southwestern bar (RA $\sim$ 11.56\degr\ and Dec $\sim$ $-$73.25\degr). 
Some `stripelike' structures are visible in the reddening map, especially for the region toward the SMC. 
This is caused by the lack of observations in the gaps between the VMC fields.

Thanks to the large size of our stellar sample, we are able to achieve
very fine resolutions and capture the fractal dust in very good detail, especially
in the regions toward the LMC and SMC centre.
The angular resolution of our resultant map is displayed in Fig.~\ref{reso}. The resolution of the reddening map 
is $\sim$ 26\,arcsec in the central parts of the LMC and SMC, and decreases down to $\sim$55\,arcmin in the MCs outskirts. 
The typical resolution of our map is $\sim$ 1.7\,arcmin. 

Our reddening maps can be accessed at \\
{\url{http://paperdata.china-vo.org/diskec/mcdust/mcmap.fits}.} \\
To use our maps, we provide a simple {\sc python} procedure in our Github project 
(\url{https://github.com/helongguo/MCdustmaps_chen2021}), which returns
reddening values for input positions  (RA and Dec). A simple example of how to use the procedure is also given in the project. 
The final catalogue containing the best-fit values of 
$E(B-V)$ from our SED fitting of $\sim$ 4 million stars, is
publicly available via: \\
{ \url{http://paperdata.china-vo.org/diskec/mcdust/starebv.fits}.}

Finally, we note that our reddening maps are two-dimensional (2D) maps, which lack the distance information. In Fig.~\ref{ebvdist} we 
plot the distributions of the reddening values of stars in a example pixel of the SMC. The distribution of reddenings
along the example line-of-sight is bimodal.
If we assume that all the dust reddening in the region comes from a relatively thin (comparing to the stellar disk) layer of dust along
the line of sight. We can thus assume that stars either experience all of
the dust column, or none of it. The probability distribution of the reddening values then becomes bimodal.
As shown in the Figure, due to the limited depths of our adopted photometric data, there are more stars in 
our sample that lie in front of the dust layer than those lie behind. Our algorithm, which
calculates the average reddening value of the pixel after excluding the outliers, tends to present the reddening values
of stars in front of the dust layer. Thus for stars embedded or behind the dust layer, values from our reddening maps may 
be underestimated.

\subsection{The MW foreground reddening maps of the area toward the MCs}\label{section4.2}

The resultant reddening maps as shown in Fig.~\ref{ebv2d} are calculated from the stars belonging to the MCs, which present
the integrated dust reddening along the lines of sight. They are important tools for the reddening correction of sources in the MCs.
However, to study the dust distribution in the LMC and SMC galaxies, we need to remove the MW foreground dust contamination.
Based on our selected MW sample stars, we have constructed the MW foreground reddening maps of the sky area toward the MCs, 
which are plotted in Fig.~\ref{mwebv}.
In the regions toward the MCs, the MW foreground dust show inhomogeneous clumpy features.
Dust clouds can be visible in the south of the LMC disk (RA $\sim$ 60\degr\ -- 105\degr\ and Dec $\sim$ $-$73\degr) 
and in the east of the SMC disk (RA $\sim$ 20\degr\ -- 30\degr\ and Dec $\sim$ $-$75\degr), where the typical
reddening value could be $E(B-V)$ $\sim$ 0.15\,mag.   
The average reddening values of the foreground MW dust is $E(B-V)$ $=$ 0.06\,mag in the LMC region and  0.04\,mag
in the area toward the SMC.

\subsection{Local dust distributions of the LMC and the SMC}

Finally, we subtract the MW foreground dust contamination (Fig.~\ref{mwebv}) from the reddening maps of the MCs
as shown in Fig.~\ref{ebv2d} and obtain the local dust 
distributions of the LMC and SMC galaxies, which are shown in Fig.~\ref{mclocal}.

Fig.~\ref{mclocal} represents the very first high angular resolution local dust reddening maps of the LMC and SMC galaxies.
Overall, the dust in the LMC and SMC are mainly located in the disk regions of the galaxies. The
previously known molecular clouds such as the 30 Dor
complex (RA = $5^{\rm{h}}38^{\rm{m}}38^{\rm{s}}$, Dec = $69^\circ5.^\prime07$), 
LMC-114 (RA = $5^{\rm{h}}23^{\rm{m}}$, Dec = $68^\circ$; \citealt{Fukui2008}) and 
LMC-154 (RA = $5^{\rm{h}}32^{\rm{m}}$, Dec = $68^\circ30^\prime$; \citealt{Fukui2008}) are clearly visible. 
The local dust maps of the LMC and SMC galaxies show very similar features to the spatial structures of the LMC 
and SMC that traced by the young stars in the previous works (e.g. 
Fig.\,A.6 and Fig.\,A.7 of \citealt{Gaia2021MCs}), which suggests the robustness of our result.

\section{Discussion}\label{section5}

\subsection{Comparison of reddening values of the individual stars from different samples}

We have selected four multi-band photometric samples (i.e. the GSV, GS2, GMV and GM2 catalogues) 
from the optical and near-IR survey catalogues and applied our SED algorithm to 
the four catalogues independently. There are common stars in the four catalogues and their reddening values have 
been calculated more than once. These common stars provide an opportunity to examine the precision of reddening values delivered 
by our SED method. We plot the distribution of the differences of reddening values deduced from the common stars from different samples 
in Fig.~\ref{intern}. There are no systematic offsets. The internal precision of our resultant reddening values, 
which can be estimated by the dispersions of $\Delta E(B-V)$ divided by a square root of 2, is about 0.04\,mag.

\subsection{Comparison of reddening values of the individual stars with previous work}

We have also compare our derived reddening values with measurements from \citet{Guo2021}. Values of $r$-band extinction of over 17 million stars 
in the southern sky are provided by \citet{Guo2021}, which are obtained
by SED fitting to photometric measurements from the SMSS DR1, 2MASS, the Wide-Field Infrared Survey Explorer 
(WISE; \citealt{Wright2010}) and Gaia DR2. 

The comparison result is shown in Fig.~\ref{guo20}.
We cross-match our results with those of \citet{Guo2021} with a matching radius of 2\,arcsec and obtain 315 common stars.
The values of $A_r$ in \citet{Guo2021} are converted to $E(B-V)$ values by $E(B - V ) = 0.43 \times A_r$ \citep{Yuan2013}.
Our measurements are in good agreement with those from \citet{Guo2021}. 
The resulted differences have a small offset of 0.027\,mag and a small rms scatter of 0.064\,mag.

\subsection{Comparisons of reddening maps with previous works}

Here we compare our reddening maps with several other recent studies, including, 
\begin{enumerate}
 \item the reddening maps from \citet{Haschke2011}, which are obtained from the $E(V-I)$ reddening values of 
red clump and RR Lyrae stars selected from the data of OGLE-III,
 \item the reddening maps from \citet{Joshi2019}, which are obtained from 
the $E(V-I)$ reddening values of classical Cepheid variables selected 
from the data of OGLE-IV, 
 \item the reddening maps from \citet{Gorski2020}, which are obtained from the 
reddening values of red clump stars selected from the data of OGLE-III, 
 \item and the reddening maps from \citet{Skowron2021}, which are obtained from the 
reddening values of red clump stars selected from the data of OGLE -IV.
\end{enumerate}

Since the angular resolution of our reddening map is higher than those of the previous works for most of the fields, we 
convert our map respectively to the same angular resolutions as the maps from the literature by 2D linear interpolations.
The maps from the literature are all converted to $E(B-V)$ using the relations recommended by the authors.
We adopt the relation $E(B-V)$ = 0.725 $\times$ $E(V-I)$ for the maps of \citet{Haschke2011},
$E(B-V)$ = 0.758 $\times$ $E(V-I)$ for the maps of \citet{Joshi2019}, 
$E(B-V)$ = 0.759 $\times$ $E(V-I)$ for the maps of \citet{Gorski2020},
and $E(B-V)$ = 0.808 $\times$ $E(V-I)$ for the maps of \citet{Skowron2021}.
The comparison results are shown Fig.~\ref{litera}. Overall, our resultant reddening maps are in good agreement with 
those of the previous works. The average differences are negligible between our maps and those from \citet{Joshi2019}
and \citet{Skowron2021}. The \citet{Haschke2011} maps are systematically slightly smaller than our results, while the 
 \citet{Gorski2020} maps are slightly larger. The dispersions of the differences are small, with rms scatter of only $\sim$ 0.03 to 0.04\,mag
for all the maps.

\section{Conclusion}\label{section6}

Based on the optical and near-IR photometries from Gaia EDR3, SMSS DR2, SMASH DR2, 2MASS PSC, VMC DR4 and VMC DR5.1, we have
obtained four different multi-band photometric catalogues: the GS2, GSV, GM2 and GMV catalogue. 
The Gaia EDR3 astrometric measurements are used to select the clean LMC, SMC and MW stellar samples. 
A SED fitting algorithm is applied to the sample stars to calculate their reddening values. 
As a result, we have obtained $E(B-V)$ reddening values of $\sim$ 1.9 million stars in the LMC, 1.5 million stars in the SMC 
and 0.6 million stars in the MW. Based on the reddening values of the LMC and SMC stars, we have constructed 
high resolution reddening map of the MCs. The map has a resolution of 
$\sim$ 26\,arcsec in the centre region of the MCs, and of $\sim$55\,arcmin in the outskirts. The typical resolution is about 1.7\,arcmin. 
Our resultant high-resolution reddening map is available online (\url{https://github.com/helongguo/MCdustmaps_chen2021}), which should be
quite useful for reddening correction of sources in the MCs.

Based on the reddening values of the MW stars, we have obtained the reddening distribution of the foreground MW dust, which are finally
subtracted to obtain the local dust distributions of the LMC and SMC galaxies. The local dust map of the LMC and SMC show very similar 
features to the spatial structures of the LMC and SMC that are traced by the young stars. It should enable us to carry out a
detailed, quantitative study of the structure and properties of dust in the two galaxies.

\section*{Acknowledgements}

This work is partially supported by the National Key R\&D Program of China No. 2019YFA0405500,
National Natural Science Foundation of China 12173034, 11803029, U1531244 and U1731308 and 
Yunnan University grant No.~C176220100007.   We acknowledge the science research
grants from the China Manned Space Project with NO.\,CMS-CSST-2021-A09, CMS-CSST-2021-A08 and CMS-CSST-2021-B03. 

This work presents results from the European Space Agency (ESA) space mission Gaia. Gaia data are being processed by the Gaia Data Processing and Analysis Consortium (DPAC). Funding for the DPAC is provided by national institutions, in particular the institutions participating in the Gaia MultiLateral Agreement (MLA). The Gaia mission website is https://www.cosmos.esa.int/gaia. The Gaia archive website is https://archives.esac.esa.int/gaia.

This publication makes use of data products from the Two Micron All Sky Survey, 
which is a joint project of the University of Massachusetts and the Infrared 
Processing and Analysis Center/California Institute of Technology, funded by 
the National Aeronautics and Space Administration and the National Science Foundation.

This work is based on data products from observations made with ESO Telescopes at the La Silla or Paranal Observatories under ESO programme ID 179.B-2003.

The national facility capability for SkyMapper has been funded through ARC LIEF grant LE130100104 from the Australian Research Council, awarded to the University of Sydney, the Australian National University, Swinburne University of Technology, the University of Queensland, the University of Western Australia, the University of Melbourne, Curtin University of Technology, Monash University and the Australian Astronomical Observatory. SkyMapper is owned and operated by The Australian National University's Research School of Astronomy and Astrophysics. The survey data were processed and provided by the SkyMapper Team at ANU. The SkyMapper node of the All-Sky Virtual Observatory (ASVO) is hosted at the National Computational Infrastructure (NCI). Development and support of the SkyMapper node of the ASVO has been funded in part by Astronomy Australia Limited (AAL) and the Australian Government through the Commonwealth's Education Investment Fund (EIF) and National Collaborative Research Infrastructure Strategy (NCRIS), particularly the National eResearch Collaboration Tools and Resources (NeCTAR) and the Australian National Data Service Projects (ANDS).

This research uses services or data provided by the Astro Data Lab at NSF's National Optical-Infrared Astronomy Research Laboratory. NOIRLab is operated by the Association of Universities for Research in Astronomy (AURA), Inc. under a cooperative agreement with the National Science Foundation. This work is based on observations at Cerro Tololo Inter-American Observatory, National Optical Astronomy Observatory (NOAO Prop. ID: 2013A-0411 and 2013B-0440; PI: Nidever), which is operated by the Association of Universities for Research in Astronomy (AURA) under a cooperative agreement with the National Science Foundation.

\section*{Data availability}

The data underlying this article are available in the article and in its online supplementary material.

\bibliographystyle{mnras}
\bibliography{mcdust}

\end{document}